\documentclass[copyright]{eptcs}
\providecommand{\event}{QAPL 2011}
\usepackage{breakurl}
\usepackage{amssymb}
\usepackage{amsmath}
\usepackage{txfonts}
\usepackage{graphicx}
\usepackage{semantic}
\usepackage{mathrsfs,comment,mathbbold}
\usepackage{color,multicol}
\usepackage{ifpdf}
\usepackage{float,proof}
\usepackage[all]{xy}
\usepackage[hyperref]{ntheorem}

\newtheorem{example}{Example}

\newcommand{\TLAMBDA}{\tilde{\lambda}}
\newcommand{\RATEZERO}{\ensuremath{\lambda_0}}
\newcommand{\PARBLANK}{\mid}
\newcommand{\ACT}{\ensuremath{\textsf Act}}
\newcommand{\PREFIX}[2]{\ensuremath{#1.#2}}
\newcommand{\NEW}[1]{\ensuremath{\nu #1}}
\newcommand{\NEWTILDE}[1]{\ensuremath{\nu \tilde{#1}}}
\newcommand{\SUM}[2]{\ensuremath{#1\,+\,#2}}
\newcommand{\CONDITION}[3]{\ensuremath{[#1]#2,#3}}
\newcommand{\ifCONDITION}[2]{\ensuremath{[#1]#2}}
\newcommand{\PAR}[2]{\ensuremath{#1\parallel#2}}

\newcommand{\SPPAR}[2]{\ensuremath{#1\parallel#2}}
\newcommand{\REC}[1][A]{\ensuremath{#1}}
\newcommand{\INPUT}[2][n]{\ensuremath{#1(#2)}}
\newcommand{\OUTPUT}[2][n]{\ensuremath{#1\langle#2\rangle}}

\newcommand{\crightarrow}{\ensuremath{\xymatrix@1@-2pt@M=1pt@!R{\ar@{->}[r]&}}}
\newcommand{\cnrightarrow}{\ensuremath{\xymatrix@1@-2pt@M=1pt@!R{\ar@{->}[r]|{/}&}}}
\newcommand{\cdashrightarrow}{\ensuremath{\xymatrix@1@-2pt@M=1pt@!R{\ar@{-->}[r]&}}}
\newcommand{\crightsquigarrow}{\ensuremath{\xymatrix@1@-2pt@M=1pt@!R{\ar@{~>}[r]&}}}
\newcommand{\nrightsquigarrow}{\ensuremath{\xymatrix@1@-2pt@M=1pt@!R{\ar@{~>}[r]|{/}&}}}
\newcommand{\dotrightarrow}{\ensuremath{\xymatrix@1@-2pt@M=1pt@!R{\ar@{.>}[r]&}}}
\newcommand{\Dotrightarrow}{\ensuremath{\xymatrix@1@-2pt@M=1pt@!R{\ar@{:>}[r]&}}}
\newcommand{\ndotrightarrow}{\ensuremath{\xymatrix@1@-2pt@M=1pt@!R{\ar@{.>}[r]|{/}&}}}
\newcommand{\nDotrightarrow}{\ensuremath{\xymatrix@1@-2pt@M=1pt@!R{\ar@{:>}[r]|{/}&}}}
\newcommand{\ndashrightarrow}{\ensuremath{\xymatrix@1@-2pt@M=1pt@!R{\ar@{-->}[r]|{/}&}}}

\newcommand{\TRAN}[3]{\ensuremath{#1 \xymatrix@1@-2pt@M=1pt@!R{\ar@{->}[r]^-{#2}&}#3}}
\newcommand{\nTRAN}[2]{\ensuremath{#1 \xymatrix@1@-2pt@M=1pt@!R{\ar@{->}[r]^-{#2}|{/}&}}}
\newcommand{\bTRAN}[3]{\ensuremath{#1 \xymatrix@1@-2pt@M=1pt@!R{\ar@{->}[r]^-{#2}&}_b#3}}
\newcommand{\TRANM}[3]{\ensuremath{#1 \xymatrix@1@-2pt@M=1pt@!R{\ar@{~>}[r]^-{#2}&} #3}}
\newcommand{\TRANI}[3]{\ensuremath{#1 \xymatrix@1@-2pt@M=1pt@!R{\ar@{-->}[r]^-{#2}&} #3}}
\newcommand{\nTRANI}[2]{\ensuremath{#1 \xymatrix@1@-2pt@M=1pt@!R{\ar@{-->}[r]|{/}^-{#2}&}}}
\newcommand{\bTRANI}[3]{\ensuremath{#1 \xymatrix@1@-2pt@M=1pt@!R{\ar@{-->}[r]^-{#2}&}_b #3}}
\newcommand{\bTRANM}[3]{\ensuremath{#1 \xymatrix@1@-2pt@M=1pt@!R{\ar@{~>}[r]^-{#2}&}_b #3}}
\newcommand{\WEAKTRANI}[3]{\ensuremath{#1 \xymatrix@1@-2pt@M=1pt@!R{\ar@{:>}[r]^-{#2}&} #3}}

\newcommand{\FN}[1]{\ensuremath{\mathit{fn}(#1)}}
\newcommand{\BN}[1]{\ensuremath{\mathit{bn}(#1)}}
\newcommand{\SUB}[2]{\ensuremath{\{#2/#1\}}}
\newcommand{\PD}[1]{\ensuremath{\mathbb{#1}}}
\newcommand{\PDA}[1]{\ensuremath{\mathbb{A#1}}}
\newcommand{\OPDA}[1]{\ensuremath{\mathcal{A}(#1)}}
\newcommand{\PREAL}{\ensuremath{\mathbb{R}_{>0}}}
\newcommand{\DEF}{\ensuremath{\stackrel{\mathit{def}}{=}}}
\newcommand{\MATRIX}{\ensuremath{\mathcal{M}}}
\newcommand{\DER}[1]{\ensuremath{\textsf{Der}(#1)}}
\newcommand{\PRATE}[1]{\ensuremath{\mu(#1)}}
\newcommand{\PINFRATE}[1]{\ensuremath{\mu_{\infty}(#1)}}

\newcommand{\MATHITSF}[1]{\ensuremath{\textsf{#1}}}
\newcommand{\iPROP}[1]{\ensuremath{\MATHITSF{Pr}^i(#1)}}

\title{A Stochastic Broadcast $\pi$-Calculus\thanks{Supported by the VKR Center of Excellence MT-LAB.}}
\author{Lei Song
\institute{Programming, Logic, and Semantics Group\\
IT University of Copenhagen, Denmark}
\email{leis@itu.dk}
\and
Flemming Nielson \qquad\qquad Bo Friis Nielsen
\institute{Department of Informatics and Mathematical Modeling\\
Technical University of Denmark}
\email{\quad nielson@imm.dtu.dk \quad\qquad bfn@imm.dtu.dk}
}
\def\titlerunning{A Stochastic Broadcast $\pi$-Calculus}
\def\authorrunning{L. Song, F. Nielson \& B.F. Nielsen}

\begin{document}
\maketitle

\begin{abstract}
In this paper we propose a stochastic broadcast $\pi$-calculus which
can be used to model server-client based systems where
synchronization is always governed by only one participant.
Therefore, there is no need to determine the joint synchronization rates.
We also take immediate transitions into account which is useful to model
behaviors with no impact on the temporal properties
of a system. Since immediate transitions may introduce non-determinism, we
will show how these non-determinism can be resolved, and as result a valid
CTMC will be obtained finally. Also some practical examples are given to show
the application of this calculus.
\end{abstract}

\section{Introduction}
Process algebras such as CCS \cite{milner1989communication}, CSP
\cite{hoare1978communicating}, and ACP \cite{bergstra1984process}
have been successfully used to model and analyze concurrent systems.
The system behavior of these classical process algebras is usually
given by \emph{labeled transition systems} (LTS) which have proved to be
a convenient framework for analyzing \emph{qualitative}
properties of large complex system. As these models are only
concerned about functional aspects of concurrent systems, process
algebras have been extended with stochastic variables in order to
model performance-oriented systems in recent years. Such examples
include TIPP \cite{gotz1993multiprocessor}, PEPA
\cite{hillston1996compositional}, EMPA \cite{bernardo1998tutorial},
stochastic $\pi$-calculus \cite{priami1995stochastic}, IMC \cite{hermanns2002interactive}, StoKlaim
\cite{de2007model}, and Stochastic Ambient Calculus \cite{vigliotti2006stochastic}. The semantics of these models are given by a
variant of LTS, \emph{Continuous Time Markov Chain} (CTMC), which
can be used to analyze \emph{quantitative} properties directly. Each
transition in a CTMC is associated with an exponentially distributed
random variable which specifies the duration of this transition. The
underlying CTMC captures the necessary information for both
functional verification and performance evaluation.

Synchronization in stochastic scenarios have been addressed in \cite{hillston1996compositional,hermanns2002interactive,aldini2009process} using different techniques. In this paper we develop a stochastic broadcast $\pi$-calculus aiming at modeling server-client based systems which are used widely in practice. In such systems synchronization are always governed by one participant, so there is
no need to determine synchronization rates like others. In
our calculus only outputs are associated with rates and their
durations are exponentially distributed while inputs are always
passive. We all know that the nondeterministic choices among outputs
can be resolved by \emph{race conditions} probabilistically.
Similarly, to resolve nondeterministic choices among inputs, we let
each input be associated with a weight as usual and the probability
of an input receiving a message is determined by its weight and the
total weight of all current inputs.  In addition the communication
in our calculus is based on broadcast, that is, when one component
outputs a message, it will be received by all the recipients instead
of only one of them. Such scenarios can be found in practice very
often. For example considering the checking out in a supermarket,
the arrivals of customers can be assumed to be exponentially
distributed. When a customer comes to the counters, he/she will
choose different counters according to the lengths of their queues,
the longer the queue the less likely it will be chosen,
meanwhile when the customer is checking out, not only the counter
knows it but also other departments will know and react accordingly
such as financial, purchasing and so on.

To enhance the expressiveness of our calculus we also take
\emph{immediate actions} into account. The immediate actions will
happen instantaneously and have been studied in
\cite{bernardo1994mpa,gotz1993multiprocessor,hermanns1995formal}.
They are useful to describe certain management and control
activities which have no impact on the temporal behavior of a
system. Since immediate action takes no time to execute, so race condition does not apply here. Instead we assign each immediate action a weight to resolve nondeterministic choices between immediate transitions which is similar as inputs. For
instance activities such as "when the buffer of a server is full,
the coming clients will be transferred to another server
instantaneously" can only be modeled by using immediate actions. In
this paper we give several classical models from performance
analysis which can be modeled in a compositional way by making use of immediate transitions. Accordingly, we will call the non-immediate transitions (resp. actions) \emph{Markovian transitions (resp. actions)}  in the sequel.

Usually the problem of immediate action is that the existence of an
underlying CTMC can no longer be guaranteed. In this paper, we solve
this in two ways. As usual immediate transitions take no time and
should have priority over Markovian transitions, so when
an immediate transition is available, it will block the executions
of Markovian transitions. We divide the whole process
space into two sets: \emph{Immediate Processes} ($\MATHITSF{IP}$) and \emph{Markovian Processes} ($\MATHITSF{MP}$).
$\MATHITSF{IP}$ only contains processes where at least one immediate
transition is available and $\MATHITSF{MP}$ contains processes where
no immediate transition is available. Since immediate transitions can exempt the execution of Markovian transitions, we can say that states in $\MATHITSF{IP}$ can only perform immediate transitions.  All states in a CTMC will belong to $\MATHITSF{MP}$. To calculate the rate from $P$ to $P'$ in a CTMC,
we accumulate the rates of all the possible transitions from $P$ to
$P'$ where transitions might be via
states in $\MATHITSF{IP}$. Sometimes it is possible for a process
reaching a state which and all its derivations belong to
$\MATHITSF{IP}$. In this case, no time is allowed to elapse and the
process is said to be \emph{absorbing}. We use a special state
$\MATHITSF{Stuck}$ to denote such situation and show how a CTMC
can be obtained even with the existence of immediate actions.

Similar with the existing calculi whose semantics are given by LTS,
we also give the LTS for our calculus. Differently, each Markovian
transition in our LTS is labeled by a rate instead of an action. For
example, a typical transition looks like
$\TRAN{P}{\lambda}{\PDA{P}}$ where $\lambda$ denotes that the execution
time of this transition is exponentially distributed with rate
$\lambda$ and $\PDA{P}$ is a distribution over pairs of action and
process $(\alpha,Q)$. Intuitively, if $\TRAN{P}{\lambda}{\PDA{P}}$,
that means that $P$ will leave its original state with rate $\lambda$ (sojourn time of $P$ is exponentially distributed with rate $\lambda$) and
get to $Q$ via action $\alpha$ with probability $p$ if the
probability of $(\alpha,Q)$ in $\PDA{P}$ is equal to $p$. By
defining an LTS in this way, the correspondent CTMC can be obtained in a
natural way. It is worth mentioning that our framework could also be
used as an alternative general way to specify the LTS as rate-base
transition systems \cite{de2009rate}. Without relying on different
techniques, for example multi relations, proved transition systems
and unique rate names used in PEPA, stochastic $\pi$-calculus, and
StoKlaim respectively, we can have a uniform way to define the
underlying models for these stochastic calculi.

The paper is organized as follows: the syntax of our calculus is
presented in the next section and in Section 3 we give the Labeled
Transition System. In Section 4 we illustrate the use of immediate
transitions by giving a few examples. We show how to get the
underlying CTMC even with existence of immediate transitions in
Section 5. Finally, we end by concluding and
describing the future work.

\section{Syntax}
Before introducing our calculus, we first give the following general
definition of probability space. A probability space is a triplet
$\mathcal{P}=(\Omega,F,\eta)$ where $\Omega$ is a set, $F$ is a
collection of subsets of $\Omega$ that includes $\Omega$ and is
closed under complement and countable union, and $\eta : F
\rightarrow [0,1]$ is a probability distribution function such that
$\eta(\Omega) = 1$ and for any collection $\{C_i\}_i$ of at most
countably many pairwise disjoint elements of $F$, $\eta(\cup_i C_i)
= \sum_i\eta(C_i)$. A probability space $(\Omega,F,\eta)$ is
discrete if $\Omega$ is countable and $F = 2^{\Omega}$, and hence
abbreviated as $(\Omega,\eta)$. Given probability spaces
$\{\mathcal{P}_i = (\Omega_i,\eta_i)\}_{i \in I}$ and weights $w_i >
0$ for each $i$ such that $\sum_{i \in I}w_i = 1$, the \emph{convex
combination} $\sum_{i \in I}w_i\mathcal{P}_i$ is defined as the
probability space $(\Omega,\eta)$ such that $\Omega = \bigcup_{i \in
I}\Omega_i$ and for each set $Y \subseteq \Omega$, $\eta(Y) =
\sum_{i \in I}w_i\eta_i(Y \cap \Omega_i)$. Usually, we use
$\{\rho_i: P_i\}_{i \in I}$ to denote a probability space
$\mathcal{P}=(\{P_i\}_{i\in I},\eta)$ such that $\eta(\{P_i\}) =
\rho_i$, here $I$ is a countable index set. \emph{Dirac} probability
space $\{1 : P\}$ will be written as $P$ directly in the sequel. If
$\sum_{i \in I}\rho_i \leq 1$ then we call it a \emph{sub
probability space}. We also use $\mathcal{P}(P_i) = \rho_i$ to
denote the probability of $P_i$ in $\mathcal{P}$. The summation and
parallel between two sub probability spaces can be defined in a
natural way as follows:
\[\SUM{\mathcal{P}_1}{\mathcal{P}_2} = \{\rho_1 + \rho_2 : P \mid \mathcal{P}_1(P) = \rho_1 \wedge \mathcal{P}_2(P) = \rho_2\wedge\mathcal{P}_1(\Omega_1)+\mathcal{P}_2(\Omega_2)\leq 1\},\]
\[\PAR{\mathcal{P}_1}{\mathcal{P}_2} = \{\rho_1 \times \rho_2 :
\PAR{P_1}{P_2} \mid \mathcal{P}_1(P_1) = \rho_1 \wedge
\mathcal{P}_2(P_2) = \rho_2\}.\]
 Note in the above $\mathcal{P}_1(\Omega_1)+\mathcal{P}_2(\Omega_2)\leq 1$ is used to guarantee that $\SUM{\mathcal{P}_1}{\mathcal{P}_2}$ is still a valid sub probability space.

We presuppose a countable set $\mathscr{C}$ of constants and a
countable set $\mathscr{V}$ of variables ranged over by $a,b,c
\ldots$ and $x,y,z\ldots$ respectively such that $\mathscr{C} \cap
\mathscr{V} = \emptyset$. $n,m,l\ldots \in \mathscr{C}\cup
\mathscr{V}$ are called names. The syntax of processes is given as follows where
$\lambda \in \PREAL$ is the exponential rate and $w \in \PREAL$ is
the weight of the input action. When the rate of an output is
infinite, it is an \emph{immediate} action which takes no time for
it to be performed. We use $\infty_w$ to denote an infinite
rate with weight $w$. In the following $\TLAMBDA$ is used to
denote either $\lambda$ or $\infty_w$ and $\RATEZERO$ ranges
over exponential rates as well as 0, that is, $\RATEZERO \in
\mathbb{R}_{\geq 0}$. It is obvious that every output action must be prefixed
by an exponential rate and every input action has a specified
weight, if the rate of an output is infinite then it will be
assigned with a weight instead. We assume that there is a countable
set of constants, ranged over by $A$, which are used to denote processes. By giving an
equation such that $A \DEF P$ we say that constant $A$ will behave
as $P$, here $A$ is required to be guarded in $P$, i.e. every constant appearing in $P$ has to be prefixed by $\ACT$. We only consider closed processes here and use $P,Q \cdots$ to range over closed processes $\mathscr{P}$.
\[\ACT ::= \INPUT{x,w} \PARBLANK \OUTPUT{m,\TLAMBDA}\]
\[P,Q ::= 0 \PARBLANK \PREFIX{\ACT}{P} \PARBLANK \NEW{a}{P} \PARBLANK \SUM{P}{Q} \PARBLANK \CONDITION{n=m}{P}{Q} \PARBLANK \PAR{P}{Q} \PARBLANK \REC\]

A substitution $\SUB{x}{a}$ can be applied to a process or process
distribution. When applied to a process distribution, it means
applying this substitution to each process with probability greater
than 0 in it. The set of free names and bound names in $P$, denoted
by $\FN{P}$ and $\BN{P}$ respectively, are defined as expected and
$\mathit{n}(P) = \FN{P} \cup \BN{P}$ denotes the set of all the
names in $P$. Structural congruence, $\equiv$, is the least
equivalence relation and congruence closed by the rules in Table
\ref{tab:structural congruence} and $\alpha$-conversion. $\equiv$ is
also extended to network distributions as usual.
\begin{table}
\centering
\caption{Structural Congruence}\label{tab:structural
congruence} \fbox{
\begin{minipage}{9cm}
\setlength{\topsep}{3 pt plus 1 pt minus 3 pt}
\setlength{\abovedisplayskip}{3 pt plus 1 pt minus 3 pt}
\setlength{\belowdisplayskip}{\abovedisplayskip}
\setlength{\abovedisplayshortskip}{1 pt plus 1 pt minus 0.5 pt}
\setlength{\belowdisplayshortskip}{1 pt plus 1 pt minus 0.5 pt}
\begin{align*}
&\NEW{a}\NEW{b}P \equiv \NEW{b}\NEW{a} P \quad \SUM{P}{Q} \equiv
\SUM{Q}{P} \quad \PAR{P}{Q} \equiv \PAR{Q}{P} \quad \CONDITION{a=a}{P}{Q} \equiv P\\
&\CONDITION{a=b}{P}{Q} \equiv Q \ a \neq b \quad \PAR{(\NEW{a}
P)}{Q} \equiv \NEW{a}(\PAR{P}{Q})\ a \notin \FN{Q}
\end{align*}
\end{minipage} }
\end{table}
\section{Semantics}\label{sec:semantics}
The actions of processes $\mathcal{A}$, ranged by $\alpha,\beta
\cdots$, are defined by
\[\alpha ::= \INPUT[a]{x} \PARBLANK \NEWTILDE{b}{\OUTPUT[a]{b}} \PARBLANK \tau\]
Here $\tilde{b}$ is a set of constants, when $b \in\tilde{b}$, $b$ is bounded, otherwise it is free. The functions
$\mathit{fn},\mathit{bn}$, and $n$ can be lifted from processes to
actions as usual.

 To evaluate the total weight of inputs on a given channel in a process, we
define function $\gamma:\mathscr{C} \times \mathscr{P} \rightarrow
\PREAL$ as Table \ref{tab:definition of weight function}.
\begin{table}[t]
\caption{Function $\gamma$ evaluating weight of input on a given
channel}\label{tab:definition of weight function}
\fbox{
\begin{minipage}{1\textwidth}
$$
\begin{array}{lcl}
\gamma(a, 0) &=& 0 \\
\gamma(a,\ACT.P) &=&
\begin{cases}
w \quad \ACT = \INPUT[a]{x,w}\text{ for some }x\\
0 \quad otherwise
\end{cases}\\
\gamma(a, \NEW{b}{P}) &=&
\begin{cases}
0 & a = b\\
\gamma(a, P) & otherwise
\end{cases}\\
\gamma(a,\SUM{P}{Q}) &=& \gamma(a,P) + \gamma(a,Q)\\
\gamma(a,\CONDITION{b=c}{P}{Q}) &=&
\begin{cases}
\gamma(a,P) \qquad b
= c\\
\gamma(a,Q) \qquad b \neq c
\end{cases}\\
\gamma(a,\PAR{P}{Q}) &=& \gamma(a,P) + \gamma(a,Q)\\
\gamma(a,\REC) &=& \gamma(a,P)\qquad A \DEF P\\
\end{array}
$$
\end{minipage} }
\end{table}

\begin{table}[t]
\caption{Function $\mu$ evaluating rate on a channel of a
process}\label{tab:definition of rate function}
\fbox{\begin{minipage}{1\textwidth}
$$
\begin{array}{lcl}
\PRATE{a,0} &=& 0 \\
\PRATE{a,\PREFIX{\OUTPUT[b]{m,\TLAMBDA}}{P}} &=& \begin{cases}\lambda & \TLAMBDA = \lambda \wedge a=b \\ 0 & \text{otherwise} \end{cases}\\
\PRATE{a,\PREFIX{\INPUT[n]{x,w}}{P}} &=& 0\\
\PRATE{a, \NEW{b}{P}} &=&
\begin{cases}
0 & a = b\\
\PRATE{a, P} & otherwise
\end{cases}\\
\PRATE{a,\SUM{P}{Q}} &=& \PRATE{a,P} + \PRATE{a,Q}\\
\PRATE{a,\CONDITION{b=c}{P}{Q}} &=&
\begin{cases}
\PRATE{a,P} \qquad b
= c\\
\PRATE{a,Q} \qquad b \neq c
\end{cases}\\
\PRATE{a,\PAR{P}{Q}} &=& \PRATE{a,P} + \PRATE{a,Q}\\
\PRATE{a,\REC} &=& \PRATE{a,P}\qquad A \DEF P\\
\end{array}
$$
\end{minipage}}
\end{table}

Similarly, we also give the function $\mu:\mathscr{C} \times
\mathscr{P} \rightarrow \PREAL$ to evaluate the total rate of
outputs on a given channel in a process which is defined in the
Table \ref{tab:definition of rate function}. To evaluate the weight
of outputs with infinity rates, we define function
$\mu_{\infty}:\mathscr{C} \times \mathscr{P} \rightarrow \PREAL$
which is the same as $\mu$ except that:
\[\PINFRATE{a,\PREFIX{\OUTPUT[b]{m,\TLAMBDA}}{P}} = \begin{cases} w\quad  \TLAMBDA = \infty_w\wedge a=b\\ 0\quad \text{otherwise}
\end{cases}\]

In addition, $\PRATE{P} = \sum_{a \in \mathscr{C}}\PRATE{a,P}$ and
$\PINFRATE{P} = \sum_{a \in \mathscr{C}}\PINFRATE{a,P}$ are used to
evaluate the total rate and total weight associated with infinite rates of outputs in a process.

We define \emph{process distribution}, ranged over by $\PD{P},
\PD{Q} \ldots$, as a probability space where $\Omega = \mathscr{P}$.
Similarly, \emph{process action distribution} can be defined as a
probability space where $\Omega = \mathcal{A} \times \mathscr{P}$.
We use $\PDA{P}, \PDA{Q} \ldots$ to range over process action
distributions. The set of all the actions in
$\PDA{P}$ is defined by $\OPDA{\PDA{P}} = \{\alpha \mid \exists P.
\PDA{P}(\alpha,P)>0\}$ while the corresponding sub process distribution of
$\alpha$ in $\PDA{P}$ is denoted by $\PDA{P}(\alpha) = \{\rho : P \mid \PDA{P}(\alpha,P) =
\rho > 0\}$. We will write $\PDA{P}$ as $(\alpha,\PD{P})$
if $\OPDA{\PDA{P}} = \{\alpha\}$ where $\PD{P} = \PDA{P}(\alpha)$.
In addition, we use $\PDA{P}(P) = \sum\limits_{\alpha \in \mathcal{A}}\PDA{P}(\alpha,P)$ to denote the total probability of $P$ in $\PDA{P}$.

We lift new operator to process action distributions in (\ref{eq:new operator on PAD}). If the channel is restricted, then the broadcast action will change to $\tau$; if the message is restricted, then the
broadcast action will be updated accordingly; otherwise the broadcast will stay unchanged while the new operator will be put on the result process.
\begin{equation}\label{eq:new operator on PAD}
\begin{aligned}
\NEW{a}\PDA{P} = &\{\rho : (\tau, \NEW{a}P) \mid
\PDA{P}(\NEWTILDE{b}\OUTPUT[a]{b},P) = \rho\}\\& \cup \{\rho :
(\NEW{a}\OUTPUT[b]{a}, P) \mid \PDA{P}(\OUTPUT[b]{a},P) = \rho
\wedge a \neq b\}\\& \cup \{\rho : (\alpha,\NEW{a}P) \mid
\PDA{P}(\alpha,P)=\rho \wedge a \notin \FN{\alpha}\}
\end{aligned}
\end{equation}

The semantics of our calculus is shown in Table \ref{tab:inference rules} where $I$ and $J$ are finite index sets. We use
$\crightsquigarrow$ to denote $\crightarrow$ or $\cdashrightarrow$,
where $\TRAN{}{\lambda}{}$ is the \emph{Markovian Transition} with rate $\lambda$, and
$\TRANI{}{w}{}$ is the \emph{Immediate Transition} with rate infinity and weight $w$. A transition
with rate 0 $\TRAN{}{0}{}$ is called a \emph{passive transition} \cite{hillston1996compositional}. All
the transitions have the form $\TRANM{P}{\RATEZERO}{\PDA{P}}$ which
means that $P$ will evolve into process $Q$ by performing action
$\alpha$ with probability $\rho$ if $\PDA{P}(\alpha,Q) = \rho$. In addition, when
it is a Markovian transition with rate $\lambda$, it means
that $P$ will leave to other states with rate $\lambda$ or the duration of the transition is exponentially distributed with rate $\lambda$. It is not
hard to see from the semantics that for a Markovian transition, all the
actions in the resulting distribution $\OPDA{\PDA{P}}$ are either outputs or $\tau$ actions, while for the passive transitions, $\OPDA{\PDA{P}}$ only contains an input action,
therefore can be written as $(\INPUT[a]{x},\PD{P})$ where $\PD{P}=\PDA{P}(\INPUT[a]{x})$. Rule (REC) means that process $\PREFIX{\INPUT[a]{x,w}}{P}$ can receive a message on channel $a$
and then evolve into $P$ with probability 1. Similarly, in rule
(mBRD) $\PREFIX{\OUTPUT[a]{b,\lambda}}{P}$ will leave to other
states with rate $\lambda$ and evolve into $P$ by broadcasting the
message $b$ on channel $a$ with probability 1, this is a Markovian
transition. If the rate of output is infinite, it should be
performed instantly. This is called immediate transition which is
shown by (iBRD). The weight associated with the infinite rate is
used to resolve nondeterministic choices as in input actions. Rule
(RES) only applies to Markovian and immediate transitions, since
$\lambda > 0$ can guarantee that the transition is not passive. The new
operator on process action distribution is defined by (\ref{eq:new operator on PAD}). By definition of $\gamma$ in Table
\ref{tab:definition of weight function}, if $\gamma(a,P) = 0$ which
means $P$ is not ready to receive messages on channel $a$, in this
case $P$ will ignore all the messages broadcasted on channel $a$.
This results in rule (LOS). Every input action is associated
with a weight which can be used to resolve
nondeterministic choices among different input actions
probabilistically. For example after receiving a message $b$ on
channel $a$,
$\SUM{\PREFIX{\INPUT[a]{x,2}}{P_1}}{\PREFIX{\INPUT[a]{x,1}}{P_2}}$
will evolve into $P_1\SUB{x}{b}$ with probability $\frac{2}{2 + 1}$
and $P_2\SUB{x}{b}$ with probability $\frac{1}{2 + 1}$. This is
shown in (SUM1). (PAR1) is straightforward since our calculus is
based on broadcast. Two parallelized processes will evolve together
after receiving a message on a certain channel. Intuitively, when we put
processes $P_1$ and $P_2$ together, the compositional process $P$
will leave to other states with rate $\lambda_1 + \lambda_2$ if the
rates of $P_1$ and $P_2$ for leaving their original states are
$\lambda_1$ and $\lambda_2$ respectively. Whether $P_1$ or $P_2$
will be executed first depends on the race condition, that is, $P_1$
will be executed before $P_2$ with probability
$\frac{\lambda_1}{\lambda_1 + \lambda_2}$ and the probability for
the other case is $\frac{\lambda_2}{\lambda_1 + \lambda_2}$. This is
captured by rules (SUM2) and (PAR2) when $\crightsquigarrow =
\crightarrow$. In (PAR2) when $\crightsquigarrow = \crightarrow$  we
also need to consider all the possible synchronization between $P$
and $Q$. For example, if $P$ can evolve into a sub process
distribution $\PDA{P}(\NEWTILDE{b_i}\OUTPUT[a_i]{b_i})$ after action
$\NEWTILDE{b_i}\OUTPUT[a_i]{b_i}$ and $Q$ will evolve into
$\PD{Q}_i\SUB{x}{b_i}$ after receiving $b_i$ on channel $a_i$, then
$\SPPAR{P}{Q}$ will evolve into sub process distribution
$\SPPAR{\PDA{P}(\NEWTILDE{b_i}\OUTPUT[a_i]{b_i})}{\PD{Q}_i\SUB{x}{b_i}}$
by performing action $\NEWTILDE{b_i}\OUTPUT[a_i]{b_i}$ after leaving
from the original state ($\NEWTILDE{b_i} \cap \FN{Q} = \emptyset$). Since $P$ and $Q$ may have several outputs available at the same time, we need to list all the possible synchronization and then add all the resulting
sub process action distributions to form the final result. The following example is to show how (PAR2) works.
\begin{example}
Suppose we have two processes: $P=\PAR{\OUTPUT[n]{y,3}}{(\SUM{\PREFIX{\INPUT[m]{x,2}}{P_1}}{\PREFIX{\INPUT[m]{x,4}}{P_2}})}$ and \\ $Q=\PAR{\OUTPUT[m]{z,2}}{\PREFIX{\INPUT[n]{x,1}}{Q_1}}$. By the semantics,
$\TRAN{P}{3}{\{1:(\OUTPUT[n]{y},\SUM{\PREFIX{\INPUT[m]{x,2}}{P_1}}{\PREFIX{\INPUT[m]{x,4}}{P_2}})\}}\equiv(\OUTPUT[n]{y},\PD{P})$ and
$\TRAN{Q}{2}{\{1:(\OUTPUT[m]{z},\PREFIX{\INPUT[m]{x,1}}{Q_1})\}}\equiv(\OUTPUT[m]{z},\PD{Q})$. When we put the two processes in parallel, we have to consider all possible synchronization between them. $P$ can broadcast $y$ on channel $n$ and $Q$ can broadcast $z$ on channel $m$, in the meanwhile $P$ can receive a message on channel $m$ and $Q$ can receive a message on channel $n$, formally,
\[\TRAN{P}{0}{\left.\begin{cases}\frac{1}{3}:&(\INPUT[m]{x},\PAR{\OUTPUT[n]{y,3}}{P_1})\\\frac{2}{3}:&(\INPUT[m]{x},\PAR{\OUTPUT[n]{y,3}}{P_2})\end{cases}\right\}\equiv(\INPUT[m]{x},\PD{P}')},\]
\[\TRAN{Q}{0}{\{1:(\INPUT[n]{x},\PAR{\OUTPUT[m]{z,2}}{Q_1})\}}\equiv(\INPUT[n]{x},\PD{Q}').\]
In $\PAR{P}{Q}$, either $P$ or $Q$ will broadcast a message first, and the non-determinism is resolved probabilistically by race condition, i.e. $\PAR{P}{Q}$ will perform $\OUTPUT[n]{y}$ first with probability $\frac{3}{5}$ and the probability of $\OUTPUT[m]{z}$ being executed first is $\frac{2}{5}$. When $\OUTPUT[n]{y}$ is performed, $Q$ will receive it and evolve into $\PD{Q}'\SUB{x}{y}$. Similarly, when $\OUTPUT[m]{z}$ is executed, $P$ will evolve into $\PD{P}'\SUB{x}{z}$ accordingly. So
\[\TRAN{\PAR{P}{Q}}{5}{\frac{3}{5}\times(\OUTPUT[n]{y},\PAR{\PD{P}}{\PD{Q}'\SUB{x}{y}}) + \frac{2}{5}\times(\OUTPUT[m]{z},\PAR{\PD{Q}}{\PD{P}'\SUB{x}{z}})}.\]
\end{example}
\begin{table}
\caption{Inference Rules($\rightsquigarrow$ denotes either
$\rightarrow$ or $\dashrightarrow$)} \label{tab:inference rules}
\small{ \fbox{
\begin{minipage}{1\textwidth}
\begin{align*}
&\infer[\text{(REC)}]{\TRAN{\PREFIX{\INPUT[a]{x,w}}{P}}{0}{\{1:(a(x),P)\}}}{}\quad
\infer[\text{(mBRD)}]{\TRAN{\PREFIX{\OUTPUT[a]{b,\lambda}}{P}}{\lambda}{\{1:(\OUTPUT[a]{b},P)\}}}{}\\
&\infer[\text{(iBRD)}]{\TRANI{\PREFIX{\OUTPUT[a]{b,\infty_{w}}}{P}}{w}{\{1:(\OUTPUT[a]{b},P)\}}}{}\quad
\infer[\text{(RES)}]{\TRANM{\NEW{a}P}{\lambda}{\NEW{a}\PDA{P}}}{\TRANM{P}{\lambda}{\PDA{P}}}\\
&\infer[,x \notin \FN{P} \text{ and } \gamma(a,P) = 0\ \text{(LOS)}]{\TRAN{P}{0}{\{1:(\INPUT[a]{x},P)\}}}{}\\
&\infer[,\gamma(a,\SUM{P}{Q})\neq0\ \text{(SUM1)}]{\TRAN{\SUM{P}{Q}}{0}{(\INPUT[a]{x},\SUM{\frac{\gamma(a,P)}{\gamma(a,\SUM{P}{Q})}\PD{P}}{\frac{\gamma(a,Q)}{\gamma(a,\SUM{P}{Q})}\PD{Q}})}}{\TRAN{P}{0}{(\INPUT[a]{x},\PD{P})} \quad \TRAN{Q}{0}{(\INPUT[a]{x},\PD{Q})}}\\
&\infer[\text{(SUM2)}]{\TRANM{\SUM{P}{Q}}{\lambda_1 +
\lambda_2}{\frac{\lambda_1}{\lambda_1 + \lambda_2}\PDA{P} +
\frac{\lambda_2}{\lambda_1 + \lambda_2}\PDA{Q}}}{\TRANM{P}{\lambda_1}{\PDA{P}} \quad
\TRANM{Q}{\lambda_2}{\PDA{Q}}}\\
&\infer[\text{(SUM3)}]{\TRANM{\SUM{P}{Q}}{\lambda}{\PDA{P}}}{\TRANM{P}{\lambda}{\PDA{P}}\quad (\crightsquigarrow = \crightarrow \wedge \PRATE{Q}=0)\vee(\crightsquigarrow = \cdashrightarrow \wedge \PINFRATE{Q}=0)}\\
&\infer[\text{(PAR1)}]{\TRAN{\PAR{P}{Q}}{0}{(\INPUT[a]{x},\PAR{\PD{P}}{\PD{Q}})}}{\TRAN{P}{0}{(\INPUT[a]{x},\PD{P})}\quad\TRAN{Q}{0}{(\INPUT[a]{x},\PD{Q})}}\\
&\infer[\text{(PAR2)}]{\TRANM{\SPPAR{P}{Q}}{\lambda_1
+ \lambda_2}{\left(\begin{aligned} &\frac{\lambda_1}{\lambda_1 +
\lambda_2}(\mathop{+}_{i\in I}(\NEWTILDE{b_i}\OUTPUT[a_i]{b_i},\SPPAR{\PDA{P}(\NEWTILDE{b_i}\OUTPUT[a_i]{b_i})}{\PD{Q}_i\SUB{x}{b_i}}) + (\tau,\SPPAR{\PDA{P}(\tau)}{Q}))\\
+ &\frac{\lambda_2}{\lambda_1 + \lambda_2}(\mathop{+}_{j \in
J}(\NEWTILDE{b_j}\OUTPUT[a_j]{b_j},\SPPAR{\PD{P}_j\SUB{x}{b_j}}{\PDA{Q}(\NEWTILDE{b_j}\OUTPUT[a_j]{b_j})})
+ (\tau,\SPPAR{P}{\PDA{Q}(\tau)}))\end{aligned}\right)}}{\left(\begin{aligned}&\TRANM{P}{\lambda_1}{\PDA{P}}\quad\mathop{\cup}\limits_{j \in J}\tilde{b_j} \cap \FN{P} = \emptyset\quad\mathop{\forall}\limits_{j \in J}\NEWTILDE{b_j}\OUTPUT[a_j]{b_j} \in
\OPDA{\PDA{Q}}.\TRAN{P}{0}{(\INPUT[a_j]{x},\PD{P}_j)}\\
&\TRANM{Q}{\lambda_2}{\PDA{Q}}\quad\mathop{\cup}\limits_{i \in I}\tilde{b_i} \cap \FN{Q} = \emptyset\quad\mathop{\forall}\limits_{i \in I} \NEWTILDE{b_i}\OUTPUT[a_i]{b_i}
\in \OPDA{\PDA{P}}.\TRAN{Q}{0}{(\INPUT[a_i]{x},\PD{Q}_i)}\end{aligned}\right)}\\
&\infer[\text{(PAR3)}]{\TRANM{\SPPAR{P}{Q}}{\lambda}{\mathop{+}\limits_{i
\in I}(\NEWTILDE{b_i}\OUTPUT[a_i]{b_i},\SPPAR{\PDA{P}(\NEWTILDE{b_i}\OUTPUT[a_i]{b_i})}{\PD{Q}_i\SUB{x}{b_i}}) + (\tau,\SPPAR{\PDA{P}(\tau)}{Q})}}{\left(\begin{aligned}&\TRANM{P}{\lambda}{\PDA{P}} \quad
\mathop{\forall}\limits_{i \in I} \NEWTILDE{b_i}\OUTPUT[a_i]{b_i}
\in\OPDA{\PDA{P}}.\TRAN{Q}{0}{(\INPUT[a_i]{x},\PD{Q}_i)}\quad\mathop{\cup}\limits_{i
\in I}\tilde{b_i} \cap \FN{Q} = \emptyset\\
 &(\crightsquigarrow = \crightarrow \wedge \PRATE{Q}=0)\vee(\crightsquigarrow = \cdashrightarrow
\wedge\PINFRATE{Q}=0)\end{aligned}\right)}\\
&\infer[,A \DEF P\ \text{(CON)}]{\TRANM{\REC}{\RATEZERO}{\PDA{P}}}{\TRANM{P}{\RATEZERO}{\PDA{P}}}\quad
\infer[\text{(STR)}]{\TRANM{P}{\RATEZERO}{\PDA{P}}}{\TRANM{P \equiv Q}{\RATEZERO}{\PDA{Q}\equiv
\PDA{P}}}
\end{align*}
\end{minipage} }}
\end{table}

Rules (SUM3) and (PAR3) are similar with rules (SUM2) and (PAR2), but they
only apply to processes where $Q$ can only have a passive transition (with label 0),
this is guaranteed by $\PRATE{Q} = 0$. (SUM3) and (PAR3) cannot be omitted since in (SUM2) and (PAR2) both $P$ and $Q$ are required to have non-passive transition, while in (SUM3) and (PAR3) only one of them has non-passive transition. The arguments for these rules when $\crightsquigarrow =
\cdashrightarrow$ are similar. Rules (CON) and (STR) are standard
and need no more comments.

From the syntax and semantics we know that the nondeterministic
choices among Markovian transitions can be resolved by a race condition
while both the nondeterministic choices among immediate outputs and
inputs can be resolved based on their weights. But still there might
be nondeterministic choices during the evolution of a process, such
as nondeterminism between passive transitions and Markovian
transitions and nondeterminism between immediate transitions and
Markovian transitions. These nondeterminism can be resolved easily
since we assume that immediate transitions can preempt other
transitions while passive transitions should not be considered when
talking about the underlying CTMC of a process. We will talk about
this in details in Section~\ref{sec:underlying CTMC}.

\begin{figure}[t]
\centering
  \setlength{\unitlength}{0.05 mm}%
  \begin{picture}(1962.6, 586.1)(0,0)
  \put(0,0){\includegraphics{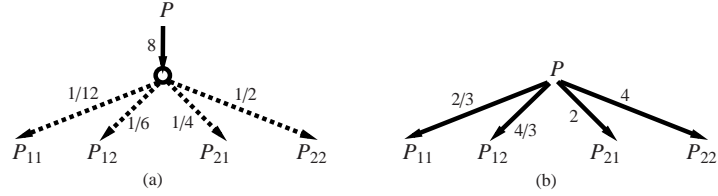}}
  \put(431.55,484.27){\fontsize{8.53}{10.24}\selectfont \makebox(90.0, 60.0)[l]{$P$\strut}}
  \put(40.00,107.28){\fontsize{8.53}{10.24}\selectfont \makebox(90.0, 60.0)[l]{$P_{11}$\strut}}
  \put(238.13,107.28){\fontsize{8.53}{10.24}\selectfont \makebox(90.0, 60.0)[l]{$P_{12}$\strut}}
  \put(537.69,107.28){\fontsize{8.53}{10.24}\selectfont \makebox(90.0, 60.0)[l]{$P_{21}$\strut}}
  \put(794.80,107.28){\fontsize{8.53}{10.24}\selectfont \makebox(90.0, 60.0)[l]{$P_{22}$\strut}}
  \put(1077.85,107.28){\fontsize{8.53}{10.24}\selectfont \makebox(90.0, 60.0)[l]{$P_{11}$\strut}}
  \put(1275.98,107.28){\fontsize{8.53}{10.24}\selectfont \makebox(90.0, 60.0)[l]{$P_{12}$\strut}}
  \put(1575.54,107.28){\fontsize{8.53}{10.24}\selectfont \makebox(90.0, 60.0)[l]{$P_{21}$\strut}}
  \put(1832.64,107.28){\fontsize{8.53}{10.24}\selectfont \makebox(90.0, 60.0)[l]{$P_{22}$\strut}}
  \put(1469.40,329.00){\fontsize{8.53}{10.24}\selectfont \makebox(30.0, 60.0)[l]{$P$\strut}}
  \put(399.57,399.19){\fontsize{6.83}{8.19}\selectfont \makebox(24.0, 48.0)[l]{8\strut}}
  \put(187.18,269.72){\fontsize{6.83}{8.19}\selectfont \makebox(96.0, 48.0)[l]{1/12\strut}}
  \put(345.22,189.51){\fontsize{6.83}{8.19}\selectfont \makebox(72.0, 48.0)[l]{1/6\strut}}
  \put(463.16,194.24){\fontsize{6.83}{8.19}\selectfont \makebox(72.0, 48.0)[l]{1/4\strut}}
  \put(630.63,267.36){\fontsize{6.83}{8.19}\selectfont \makebox(72.0, 48.0)[l]{1/2\strut}}
  \put(1208.52,253.20){\fontsize{6.83}{8.19}\selectfont \makebox(72.0, 48.0)[l]{2/3\strut}}
  \put(1375.99,173.01){\fontsize{6.83}{8.19}\selectfont \makebox(72.0, 48.0)[l]{4/3\strut}}
  \put(1522.23,208.38){\fontsize{6.83}{8.19}\selectfont \makebox(24.0, 48.0)[l]{2\strut}}
  \put(1659.04,264.99){\fontsize{6.83}{8.19}\selectfont \makebox(24.0, 48.0)[l]{4\strut}}
  \put(387.68,40.92){\fontsize{6.83}{8.19}\selectfont \makebox(72.0, 48.0)[l]{(a)\strut}}
  \put(1432.60,38.56){\fontsize{6.83}{8.19}\selectfont \makebox(72.0, 48.0)[l]{(b)\strut}}
  \end{picture}%
\caption{\label{fig:CTMCexample}A Simple CTMC}
\end{figure}

In this section we will not discuss immediate transitions, we leave
it to the next section. The following simple example is to show how
to get a \textsf{CTMC} from a process without immediate actions and
we often omit the tail process $0$.
\begin{example}
Given a process $P\equiv\PAR{\OUTPUT[a]{b_1,2}}{\PAR{\OUTPUT[a]{b_2,6}}{(\SUM{\PREFIX{\INPUT[a]{x,1}}{P_1}}{\PREFIX{\INPUT[a]{x,2}}{P_2}})}}$,
then $P$ will broadcast a message ($b_1$ or $b_2$) on channel $a$
with an exponential delay 8 and then evolve into a process action
distribution $\PDA{P}$. By semantics in Table \ref{tab:inference
rules}, we know \[\PDA{P}=\{\frac{1}{12}:(\OUTPUT[a]{b_1},P_{11}),\frac{1}{6}:(\OUTPUT[a]{b_1},P_{12}),\frac{1}{4}:(\OUTPUT[a]{b_2},P_{21}),\frac{1}{2}: (\OUTPUT[a]{b_2},P_{22})\}\]
where
\begin{align*}
P_{11} = \PAR{\OUTPUT[a]{b_2,6}}{P_1\SUB{x}{b_1}} \quad P_{12} =
\PAR{\OUTPUT[a]{b_2,6}}{P_2\SUB{x}{b_1}}\\
P_{21} = \PAR{\OUTPUT[a]{b_1,2}}{P_1\SUB{x}{b_2}} \quad P_{22} =
\PAR{\OUTPUT[a]{b_1,2}}{P_2\SUB{x}{b_2}}
\end{align*}
This is displayed in Fig.~\ref{fig:CTMCexample}(a) and the correspondent \textsf{CTMC} is
shown in Fig. \ref{fig:CTMCexample}(b), here we use dot lines to
denote probabilistic choices and omit actions of the transitions.
\end{example}

In the above example we briefly illustrated how to get a CTMC from a process. Now we are going to give the general construction by which we can
get the correspondent CTMC from a process. Use $\MATRIX(P,Q)$ to denote the rate from $P$ to $Q$ in a CTMC,
 and define $\DER{P}$ as the smallest set of processes satisfying: i) $P
\in \DER{P}$; ii) $P_2 \in \DER{P}$ iff there exists $P_1 \in
\DER{P}$ such that $\TRAN{P_1}{\lambda}{\PDA{P}}$ with
$\PDA{P}(P_2) > 0$. So $\DER{P}$ is the set of all the processes which are reachable from
$P$ with positive probability via arbitrary steps. For each two processes $P_1,P_2 \in
\DER{P}$, the rate from $P_1$ to $P_2$ is equal to $\lambda \times
\PDA{P}(P_2)$, that is $\MATRIX(P_1,P_2) = \lambda \times
\PDA{P}(P_2)$ such that $\TRAN{P_1}{\lambda}{\PDA{P}}$, otherwise
$\MATRIX(P_1,P_2) = 0$.

\section{Immediate Transitions}\label{sec:immediate actions}
In this section, we will give a few examples and show how can we
benefit from immediate transitions.

First we consider a model called \emph{Closed Queueing Networks}
(CQN) \cite{buzen1973computational} from performance analysis. A
queueing network is a collection of servers. Customers must proceed
from one server to another in order to satisfy their service
requirements. The queueing network is closed if neither arrivals nor
departures of customers are permitted; instead the number of
customers in the network are fixed at all times. We use $M$ to
denote the number of servers in the network and $N$ for the number
of customers circulating around. The service time for a customer at server $i$ is
exponentially distributed with rate $\lambda_i$ and the probability
a customer will proceed to the $j$-th server after completing a
service  request at server $i$ is equal to $p_{ij}$ for $i,j =
1,2,\ldots,M$.
\begin{example}\label{ex:CQN}
Suppose we are given a CQN with 5 servers and 15 customers shown in
Fig.~\ref{fig:CQN} where the numbers in the rectangles denote the
length of the queue of each server as well as their indexes, the numbers
on the edges denote the transition probabilities and the numbers in
the circles denote the rate of each service time.

\begin{figure}[t]
\begin{minipage}[t]{0.4\linewidth}
\centering
  \setlength{\unitlength}{0.05 mm}%
  \begin{picture}(689.0, 847.5)(0,0)
  \put(0,0){\includegraphics{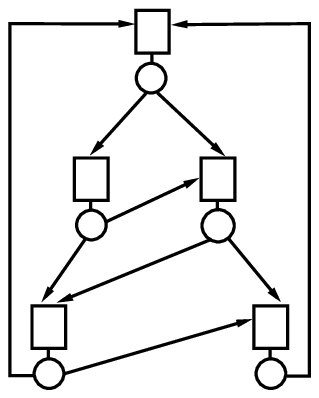}}
  \put(319.72,744.31){\fontsize{6.83}{8.19}\selectfont \makebox(24.0, 48.0)[l]{1\strut}}
  \put(195.37,437.85){\fontsize{6.83}{8.19}\selectfont \makebox(24.0, 48.0)[l]{2\strut}}
  \put(450.71,440.98){\fontsize{6.83}{8.19}\selectfont \makebox(24.0, 48.0)[l]{3\strut}}
  \put(107.57,136.13){\fontsize{6.83}{8.19}\selectfont \makebox(24.0, 48.0)[l]{4\strut}}
  \put(208.08,580.18){\fontsize{5.12}{6.15}\selectfont \makebox(54.0, 36.0)[l]{0.3\strut}}
  \put(420.89,575.90){\fontsize{5.12}{6.15}\selectfont \makebox(54.0, 36.0)[l]{0.7\strut}}
  \put(302.83,645.62){\fontsize{6.83}{8.19}\selectfont \makebox(48.0, 48.0)[l]{10\strut}}
  \put(564.00,141.38){\fontsize{6.83}{8.19}\selectfont \makebox(24.0, 48.0)[l]{5\strut}}
  \put(191.79,344.97){\fontsize{6.83}{8.19}\selectfont \makebox(24.0, 48.0)[l]{4\strut}}
  \put(440.20,345.06){\fontsize{6.83}{8.19}\selectfont \makebox(48.0, 48.0)[l]{10\strut}}
  \put(109.70,40.50){\fontsize{6.83}{8.19}\selectfont \makebox(24.0, 48.0)[l]{5\strut}}
  \put(561.15,43.44){\fontsize{6.83}{8.19}\selectfont \makebox(24.0, 48.0)[l]{3\strut}}
  \put(95.05,285.68){\fontsize{5.12}{6.15}\selectfont \makebox(54.0, 36.0)[l]{0.5\strut}}
  \put(296.12,438.54){\fontsize{5.12}{6.15}\selectfont \makebox(54.0, 36.0)[l]{0.5\strut}}
  \put(301.10,234.46){\fontsize{5.12}{6.15}\selectfont \makebox(54.0, 36.0)[l]{0.4\strut}}
  \put(474.40,230.27){\fontsize{5.12}{6.15}\selectfont \makebox(54.0, 36.0)[l]{0.6\strut}}
  \put(43.60,465.29){\fontsize{5.12}{6.15}\selectfont \makebox(54.0, 36.0)[l]{0.2\strut}}
  \put(365.64,83.87){\fontsize{5.12}{6.15}\selectfont \makebox(54.0, 36.0)[l]{0.8\strut}}
  \put(613.17,424.83){\fontsize{5.12}{6.15}\selectfont \makebox(18.0, 36.0)[l]{1\strut}}
  \end{picture}%
\caption{\label{fig:CQN}A CQN with 5 Servers}
\end{minipage}%
\begin{minipage}[t]{0.6\linewidth}
\centering
  \setlength{\unitlength}{0.05 mm}%
  \begin{picture}(948.1, 1010.6)(0,0)
  \put(0,0){\includegraphics{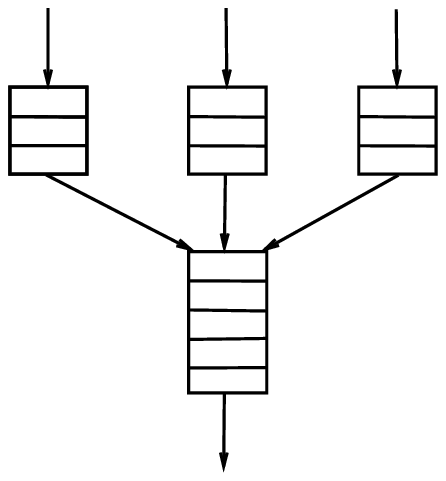}}
  \put(79.68,867.14){\fontsize{5.69}{6.83}\selectfont \makebox(20.0, 40.0)[l]{2\strut}}
  \put(445.11,868.80){\fontsize{5.69}{6.83}\selectfont \makebox(20.0, 40.0)[l]{4\strut}}
  \put(784.07,867.14){\fontsize{5.69}{6.83}\selectfont \makebox(20.0, 40.0)[l]{5\strut}}
  \put(230.15,503.37){\fontsize{5.69}{6.83}\selectfont \makebox(20.0, 40.0)[l]{3\strut}}
  \put(431.88,536.44){\fontsize{5.69}{6.83}\selectfont \makebox(20.0, 40.0)[l]{8\strut}}
  \put(732.82,521.56){\fontsize{5.69}{6.83}\selectfont \makebox(20.0, 40.0)[l]{6\strut}}
  \put(425.27,106.53){\fontsize{5.69}{6.83}\selectfont \makebox(40.0, 40.0)[l]{10\strut}}
  \put(90.96,696.35){\fontsize{7.11}{8.54}\selectfont \makebox(75.0, 50.0)[l]{$Q_1$\strut}}
  \put(453.53,696.35){\fontsize{7.11}{8.54}\selectfont \makebox(75.0, 50.0)[l]{$Q_2$\strut}}
  \put(801.52,696.35){\fontsize{7.11}{8.54}\selectfont \makebox(75.0, 50.0)[l]{$Q_3$\strut}}
  \put(455.94,304.47){\fontsize{7.11}{8.54}\selectfont \makebox(75.0, 50.0)[l]{$Q_0$\strut}}
  \end{picture}%
\caption{\label{fig:OQN with Blocking}Open Queue Network with Blocking}
\end{minipage}
\end{figure}

The communication in the CQN is "point-to-point" in the sense that
each leaving customer will arrive at only one server, while in our
calculus the communication is based on broadcast. But with immediate
transitions we can model such "point-to-point" communication as
follows. Here we assume that the weight of each input is 1 by
default, that is, $\INPUT[c_i]{x}$ is equal to $\INPUT[c_i]{x,1}$.
$\MATHITSF{SQ}_i(w_i)$ denotes server $i$ with $w_i$ customers
waiting for service in its queue. In Table~\ref{tab:CQN}, $\MATHITSF{CQN}(1,2,3,4,5)$
denotes the system where the $i$-th parameter is the length of the $i$-th queue.
\begin{table}[H]
 \caption{Model of Closed Queueing Network}\label{tab:CQN}
  \small{
  \fbox{
\begin{minipage}{0.9\textwidth}
\begin{align*}
&\MATHITSF{Rev}(i,j,w_j)=\PREFIX{\INPUT[c_i]{x}}{[x=w](\SUM{\PREFIX{\OUTPUT[c]{r,\infty_{p_{ij}}}}{(\PAR{\MATHITSF{SQ}_j(w_j+1)}{\MATHITSF{SQ}_i(w-1)})}}{\PREFIX{\INPUT[c]{x}}{\MATHITSF{SQ}_j(w_j)}})}\quad w\in[1,15]\\
&\MATHITSF{SQ}_j(w_j)=\SUM{\sum_{1\leq i\leq
5,p_{ij}>0}\MATHITSF{Rev}(i,j,w_j)}{(\CONDITION{w_j=0}{0}{\OUTPUT[c_j]{w_j,\lambda_j}})}\\
&\MATHITSF{CQN}(1,2,3,4,5)=\NEW{c}(\parallel_{i=1}^5\MATHITSF{SQ}_i(i))
\end{align*}
\end{minipage}
}}
\end{table}

Each $\MATHITSF{SQ}_j(w_j)$ contains two parts: receivers denoted by $\sum_{1\leq i\leq
5,p_{ij}>0}\MATHITSF{Rev}(i,j,w_j)$ listen on the channels of their predecessors, and after being notified that a customer is coming, it will try to broadcast on channel $c$ immediately with a specific weight $\infty_{p_{ij}}$. The one which succeeds to do so will be the destination of the customer and all the others will be informed by receiving a message on $c$. The other part $\CONDITION{w_j=0}{0}{\OUTPUT[c_j]{w_j,\lambda_j}}$ takes care of the requests of the customers in its queue if it is not empty, the rate of the customer leaving depends on the service rate. By putting these five servers in parallel, we get the whole system $\MATHITSF{CQN}(1,2,3,4,5)$. Our semantics
guarantees that each leaving customer will finally arrive at one and only one server.
Actually, this can be seen as a way to model "point-to-point"
communication with immediate transitions. For example if a customer
is leaving server 3, then both server 4 and 5 will be informed by
broadcasting a message on channel $c_3$. But after that server 4 and
5 will try to broadcast an acknowledge $r$ on channel $c$, with
probability 0.4 and 0.6 respectively. The one which succeeds to send
$r$ will be the real destination of the customer. The other server
will know this fact by listening on channel $c$ and roll back to its
original state at the same time. The model in Table~\ref{tab:CQN} is
quite flexible, for example, we can add self loops easily, that is,
the destination of a leaving customer can be the same server as its
departure server, or instead of having fixed transfer probabilities,
we can make them change based on the current lengths of queues.
\end{example}

In Example \ref{ex:CQN} we have shown a typical closed queueing network
where we assume that every server has a queue with infinite
capacity. But in practice the capacity of queues is often limited.
In \cite{van1988jackson} a variant of closed queueing network is
proposed where the queue of each server $i$ only has finite capacity
$B_i$ for $1 \leq i \leq M$. A customer which requests service at
server $i$ while the queue of server $i$ is full will instantly be
routed to another server $j$ with probability $p_{ij}$ as if it is
served by server $i$ at an infinite speed. For this kind of model,
it is hard (if not impossible) to model it in a compositional way
without using immediate actions. With immediate actions, the model
is easy to obtain without changing the model in Table~\ref{tab:CQN} a
lot.

\begin{example}\label{ex:CQN with finite queue}
Suppose the queueing network we are about to model is the same as
the queueing network in Example \ref{ex:CQN} except that the
capacity of every queue is maximum 10. If one server is full, it
will just transfer the coming customers to its next servers. The
model can be obtained by simply replacing the $\MATHITSF{SQ}_j(w_j+1)$ in
$\MATHITSF{Rev}(i,j,w_j)$ in Table~\ref{tab:CQN} with $\CONDITION{w_i=10}{\OUTPUT[c_i]{11,\infty_1}}{\MATHITSF{SQ}_j(w_j+1)}$.
This means instead of accepting any coming customers, we require an
extra checking on the current queue. If a customer arrives at a
server whose queue is not full, the customer will be accepted,
otherwise, the customer will also be accepted but will be transferred to the next server via
action $\OUTPUT[c_i]{11,\infty_1}$ just like it is served with
infinite rate, that gives excuse of parameter 11. The same process will continue until the customer
arrives at some server which has free space for it.
\end{example}
From the semantics in Table~\ref{tab:inference rules}, we know that
all the outputs are non-blocking, i.e., for one message to be
broadcasted, it is not necessary to have recipients. But sometimes
we may have models where components are not completely independent and
one component can do something only after some other components
finish, that is, some behaviors are blocking. Blocking here means
that an output action cannot happen spontaneously but has to wait
until certain conditions are fulfilled. For instance in Example
\ref{ex:CQN with finite queue}, every queue has finite capacity.
When a customer arrives at a server without free space, it will
simply be transferred to other servers. What if the customer cannot
be transferred but can only wait until the server has free spaces?
Refer to the following example from \cite{lee1989approximate} which
is also a variant of queueing network called \emph{open queueing
network with blocking}.

The network consists of $N$ parallel servers called \emph{merging} queues; there is a queue receiving the outputs of these merging queues and is called \emph{merged} queue (or queue 0). The service
time at queue $i$ is exponentially distributed with rate
$\lambda_i$. The queue network is open since the number
of the customers circulating in the network is not fixed and some
external customers may arrive from the outside. Arrivals to queue
$i$ are independent \emph{Poisson Processes} with rate $\mu_i$.
There is no external arrival to the merged queue. The length of the
$i$-th merging queue is $B_i$. The capacity of the queue 0 is $B_0$,
and $\lambda_0$ is its service rate at queue 0. If a customer
arrives at a merging queue when it is full, the customer will be
lost. When a customer completes service at server $i$, it will be transferred
to queue 0 only if it is not full; otherwise, the customer
waits in the $i$-th server until it can enter queue 0. During this
time the $i$-th server cannot serve other customers that might wait in its queue. In this case, the queue is said to be \emph{blocked} and queue 0 is \emph{blocking}. Since there are $M$
servers in parallel, there might be more than one queue blocked at the same time. When more than one queues are blocked, it is assumed that
they will enter queue 0 on a "First-Blocked-First-Enter" basis.

From the description of open queue network with blocking, we know
there are two kinds of actions involving in this model: one is
blocking and the other one is non-blocking. For instance, when queue
0 is full, any other arrivals have to wait until queue 0 has free
space, this is blocking action. On the other hand, when the merging
queues are full, instead of blocking the external arrivals it will
just discard them, so the external arrivals are non-blocking
actions. In the following we will show how to model both kinds of
actions in our calculus.

\begin{figure}[t]
\centering
  \setlength{\unitlength}{0.05 mm}%
  \begin{picture}(1615.9, 778.9)(0,0)
  \put(0,0){\includegraphics{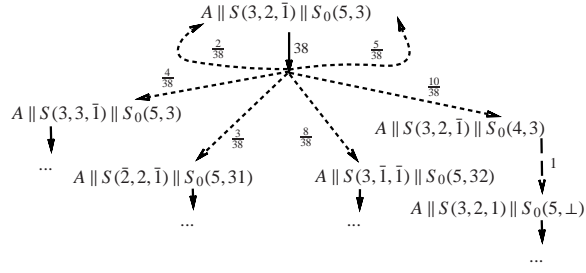}}
  \put(543.46,687.36){\fontsize{7.11}{8.54}\selectfont \makebox(525.0, 50.0)[l]{$\PAR{A}{\PAR{S(3,2,\bar{1})}{S_0(5,3)}}$\strut}}
  \put(40.00,418.91){\fontsize{7.11}{8.54}\selectfont \makebox(525.0, 50.0)[l]{$\PAR{A}{\PAR{S(3,3,\bar{1})}{S_0(5,3)}}$\strut}}
  \put(204.63,249.27){\fontsize{7.11}{8.54}\selectfont \makebox(575.0, 50.0)[l]{$\PAR{A}{\PAR{S(\bar{2},2,\bar{1})}{S_0(5,31)}}$\strut}}
  \put(846.99,252.96){\fontsize{7.11}{8.54}\selectfont \makebox(550.0, 50.0)[l]{$\PAR{A}{\PAR{S(3,\bar{1},\bar{1})}{S_0(5,32)}}$\strut}}
  \put(1006.12,380.45){\fontsize{7.11}{8.54}\selectfont \makebox(525.0, 50.0)[l]{$\PAR{A}{\PAR{S(3,2,\bar{1})}{S_0(4,3)}}$\strut}}
  \put(1100.91,167.61){\fontsize{7.11}{8.54}\selectfont \makebox(475.0, 50.0)[l]{$\PAR{A}{\PAR{S(3,2,1)}{S_0(5,\bot)}}$\strut}}
  \put(1415.05,38.50){\fontsize{7.11}{8.54}\selectfont \makebox(75.0, 50.0)[l]{...\strut}}
  \put(930.04,135.88){\fontsize{7.11}{8.54}\selectfont \makebox(75.0, 50.0)[l]{...\strut}}
  \put(487.20,135.87){\fontsize{7.11}{8.54}\selectfont \makebox(75.0, 50.0)[l]{...\strut}}
  \put(109.86,283.07){\fontsize{7.11}{8.54}\selectfont \makebox(75.0, 50.0)[l]{...\strut}}
  \put(788.79,597.05){\fontsize{5.69}{6.83}\selectfont \makebox(40.0, 40.0)[l]{38\strut}}
  \put(430.13,522.65){\fontsize{5.69}{6.83}\selectfont \makebox(80.0, 40.0)[l]{$\frac{4}{38}$\strut}}
  \put(616.68,360.29){\fontsize{5.69}{6.83}\selectfont \makebox(80.0, 40.0)[l]{$\frac{3}{38}$\strut}}
  \put(798.15,364.87){\fontsize{5.69}{6.83}\selectfont \makebox(80.0, 40.0)[l]{$\frac{8}{38}$\strut}}
  \put(1138.12,498.58){\fontsize{5.69}{6.83}\selectfont \makebox(100.0, 40.0)[l]{$\frac{10}{38}$\strut}}
  \put(1469.40,292.04){\fontsize{5.69}{6.83}\selectfont \makebox(20.0, 40.0)[l]{1\strut}}
  \put(565.03,598.45){\fontsize{5.69}{6.83}\selectfont \makebox(80.0, 40.0)[l]{$\frac{2}{38}$\strut}}
  \put(985.43,596.60){\fontsize{5.69}{6.83}\selectfont \makebox(80.0, 40.0)[l]{$\frac{5}{38}$\strut}}
  \end{picture}%
\caption{\label{fig:Fragment of OQN}Execution Fragment of $P$}
\end{figure}

\begin{example}\label{ex:OQN with blocking}
Fig.~\ref{fig:OQN with Blocking} gives a concrete example with 3
servers marked as $Q_1$, $Q_2$, and $Q_3$ respectively. The length
of each queue is 3 and the numbers on the in-edges and out-edges are
used to denote arrival rates and service rates, that is, $\lambda_1
= 3$, $\lambda_2 = 8$, $\lambda_3=6$, $\mu_1=2$, $\mu_2=4$, and
$\mu_3=5$, and $\lambda_0 = 10$. The $Q_0$ at the bottom is the queue 0
with service rate 10, and the capacity of its queue is 5. Initially, every queue is empty.

Here is the model of the queueing network in Fig.~\ref{fig:OQN with
Blocking} where $A_i$ denotes the arrival process of queue $i$,
$Q_i(w)$ denotes queue $i$ with $w$ customers in the queue for $1
\leq i \leq 3$, and $Q_0(w,\tilde{S})$ denotes queue 0 with $w$
customers. The $\tilde{S}$ is a sequence of queues which are waiting for
queue 0 when it is full, it is an element of $\MATHITSF{SQ}$ which is
defined by enumerating all the possible sequences of queues waiting
for queue 0. The symbol $\bot$ is used to denote empty sequence.
\[\MATHITSF{SQ} = \{\bot,1,2,3,12,21,13,31,23,32,123,132,213,231,312,321\}\]
For simplicity, we define two functions on this set:
$\mathit{H(\tilde{S})}:\MATHITSF{SQ}\rightarrow \{\bot,1,2,3\}$ and
$\mathit{T(\tilde{S})}:\MATHITSF{SQ} \rightarrow \MATHITSF{SQ}$
which return the head of sequence and the left sequence by deleting
the first element respectively. For example, $\mathit{H}(123) = 1$
and $\mathit{T}(123)=23$; $\mathit{H}(\bot) = \bot$ and
$\mathit{T}(\bot)=\bot$. We use $\tilde{S}i$ to denote a new sequence by attaching $i$ to the end
of $\tilde{S}$. Note here that these functions are just used to give
a compact model, they can be replaced by the standard operators by
enumerating all the possible cases. It is similar for conditions
like $[w=0,1,2 \wedge \mathit{H}(\tilde{S})=i]$.
\begin{table}
 \caption{Model of Open Queueing Network with Blocking}\label{tab:OQN with blocking}
  \small{
  \fbox{
\begin{minipage}{0.9\textwidth}
\begin{align*}
A_i &=\PREFIX{\OUTPUT[a_i]{\mathit{arrival},\mu_i}}{A_i}\\
S_i(w_i) &=
\SUM{\ifCONDITION{w_i=0,1,2}{\PREFIX{\INPUT[a_i]{x,1}}{S_i(w_i+1)}}}{\ifCONDITION{w_i=1,2,3}{\PREFIX{\OUTPUT[s_i]{\mathit{leave}_i,\lambda_i}}{\PREFIX{\INPUT[b_i]{x,1}}{S_i(w_i-1)}}}}\\
S_0(w,\tilde{S})&=\SUM{\SUM{\ifCONDITION{w=1,2,3,4,5}{\PREFIX{\OUTPUT[s_0]{\mathit{leave}_0,\lambda_0}}{S_0(w-1,\tilde{S})}}}{\ifCONDITION{w=5}{\sum_i\PREFIX{\INPUT[s_i]{x,1}}{S_0(w,\tilde{S}i)}}}\\&}{\SUM{\ifCONDITION{w=0,1,2,3,4\wedge\mathit{H}(\tilde{S})=i}{\PREFIX{\OUTPUT[b_i]{\mathit{unblocking,\infty_1}}}{S_0(w+1,\mathit{T}(\tilde{S}))}}\\&}{\ifCONDITION{w=0,1,2,3,4\wedge\mathit{H}(\tilde{S})=\bot}{\sum_i\PREFIX{\INPUT[s_i]{x,1}}{\PREFIX{\OUTPUT[b_i]{\mathit{unblocking},\infty_1}}{S_0(w+1,\tilde{S})}}}}}
\end{align*}
\end{minipage}
}}
\end{table}
The model of each component in Fig.~\ref{fig:OQN with Blocking} is
shown in Table~\ref{tab:OQN with blocking}. The whole system can be
denoted as $P = \mathop{\parallel}\limits_{i}A_i
\mathop{\parallel}\limits_{i}S_i(0) \parallel S_0(0,\bot)$ with
$1\leq i \leq 3$.

As we said before, broadcasts such that
$\OUTPUT[s_i]{\mathit{leave}_i,\lambda_i}$ are blocking, so when
there is no input $\INPUT[s_i]{x,w}$ available, that is, the queue 0
is full, server $i$ has to wait before it can perform other actions.
To do so, we let the server $i$ wait for the message
$\mathit{unblocking}$ on channel $b_i$ after one customer leaving
from it to server 0. If queue 0 has free spaces, it will perform
action $\OUTPUT[b_i]{\mathit{unblocking,\infty_1}}$ right after it
receives the request from server $i$. Otherwise if the queue 0 is
full, it will attach the request to the end of its waiting list.
When server 0 is ready to handle the request after several steps, it
will send the message $\mathit{unblocking}$ to server $i$ instantly
via immediate action $\OUTPUT[b_i]{\mathit{unblocking,\infty_1}}$.
Server $i$ will receive it at the same time and then unblock itself.

Fig.~\ref{fig:Fragment of OQN} shows a fragment of the execution of
$P$ where $\crightarrow$, $\cdashrightarrow$, and $\dotrightarrow$ denote
Markovian, immediate and probabilistic transition respectively. In
additional $A=\parallel_{i}A_i$ and $S(q_1,q_2,q_3)=\parallel_{i}S_i(q_i)$,
when $q_i$ is barred, it means that server $i$ is blocked. For example, $S(q_1,\bar{q_2},q_3) =
\PAR{S_1(q_1)}{\PAR{\PREFIX{\INPUT[b_2]{x,1}}{S_2(q_2)}}{S_3(q_3)}}$.
When in state $\PAR{A}{\PAR{S(3,2,\bar{1})}{S_0(4,3)}}$, it means
that there is a free space in queue 0 while the server 3 is waiting
for service. In this case, server 0 should response to it and
transfer to the state $\PAR{A}{\PAR{S(3,2,1)}{S_0(5,\bot)}}$
instantly where the request is removed from the waiting list to the
queue 0. The self loop with probability $\frac{2}{38}$ denotes that
the arrivals of external customers to server 1 while its queue is
full. In this case, the arriving customers will be discarded without
causing any effects. The other self loop with probability
$\frac{5}{38}$ is similar except that the arriving customer is
discarded because server 3 is blocked.
\end{example}

From Example~\ref{ex:OQN with blocking}, we can see that blocking
actions can be represented easily by using immediate transitions. In
general when broadcast $\OUTPUT[a]{b,\lambda}$ is blocking, it
should be prefixed with an input such as
$\PREFIX{\INPUT[c]{x,w}}{\OUTPUT[a]{b,\lambda}}$. When certain
conditions are fulfilled, the process should trigger
$\OUTPUT[a]{b,\lambda}$ by sending a message on channel $c$
instantly, that is, by action like
$\OUTPUT[c]{\mathit{unblocking},\infty_w}$.

\section{The Underlying CTMC}\label{sec:underlying CTMC}
In Examples~\ref{ex:CQN}, \ref{ex:CQN with finite queue}, and
\ref{ex:OQN with blocking}, we see that immediate transition is
indeed powerful to model some systems. But the main disadvantage is
that the underlying CTMC of a process is not so obvious anymore. In
this section we will show how to define the underlying CTMC in case
 of immediate transitions. Different from \cite{bernardo2007weak} where the eliminations of immediate transitions are based on the weak behavioral equivalence, we solve this by dealing with a set of equations as follows.

In this calculus choices between Markovian actions are probabilistic
depending on their rates while choices between immediate actions are
also probabilistic depending on their associated weights. In
addition, if both types of actions are involved in a choice, the
immediate action should be prioritized, since they take no time, so
the probability of the Markovian action being executed before the
immediate one is zero \cite{hermanns1995formal}.  Since the priority
of immediate actions are not shown in Table \ref{tab:inference
rules}, a CTMC cannot be obtained directly from a process based on the
semantics. In the following we distinguish between $\MATHITSF{MP}$ and $\MATHITSF{IP}$.
$\MATHITSF{MP}$ only contains processes which do
not have immediate transitions while $\MATHITSF{IP}$ only contains
processes which have immediate transitions available. Formally,
$\MATHITSF{IP} = \{P \in \mathscr{P} \mid \exists
\PDA{P}.\TRANI{P}{\lambda}{\PDA{P}}\}$ and $\MATHITSF{MP} = \{P
\in \mathscr{P} \mid
\nexists\PDA{P}.\TRANI{P}{\lambda}{\PDA{P}}\}= \mathscr{P}
\setminus \MATHITSF{IP}$. It is not hard to see that every state in
a CTMC should be seen as a Markovian process in this calculus, so
instead of considering all the processes in $\DER{P}$ as in Section~\ref{sec:semantics}, we
only need to consider set $\DER{P} \cap \MATHITSF{MP}$ when talking
about the corresponding CTMC of $P$. The question now is how to give
the value of $\MATRIX(P,Q)$ for any $P,Q \in \DER{P} \cap
\MATHITSF{MP}$. Due to immediate actions, it is not enough to just
consider one step Markovian transition as before since $P$ might
have to go through several immediate processes before reaching $Q$.

We use $\iPROP{R,Q}$ to denote the probability from $R$ to $Q$ via
all possible immediate transitions where $R \in \mathscr{P}$ and $Q
\in \MATHITSF{MP}$.  The value of $\iPROP{R,Q}$ is given by the
smallest solution defined by the following set of equations:
\begin{equation}\label{eq:prob}\iPROP{R,Q} =
\begin{cases}1 & R=Q\\
\sum_{\PDA{R}(P)>0}\PDA{R}(P)\times\iPROP{P,Q}  &
\TRANI{R}{\lambda}{\PDA{R}} \wedge \DER{R} \cap \MATHITSF{MP}\neq\emptyset\\
0 &\text{otherwise}
\end{cases}\end{equation}
Then for any $P,Q \in \MATHITSF{MP}$, $\MATRIX(P,Q)$ can be defined
as follows:
\[\MATRIX(P,Q)=\lambda\times\sum_{\PDA{P}(R)>0}\PDA{P}(R)\times\iPROP{R,Q}\quad \TRAN{P}{\lambda}{\PDA{P}}\]

\begin{figure}[t]
\begin{minipage}[t]{0.6\textwidth}
  \setlength{\unitlength}{0.05 mm}%
  \begin{picture}(1530.3, 968.6)(0,0)
  \put(0,0){\includegraphics{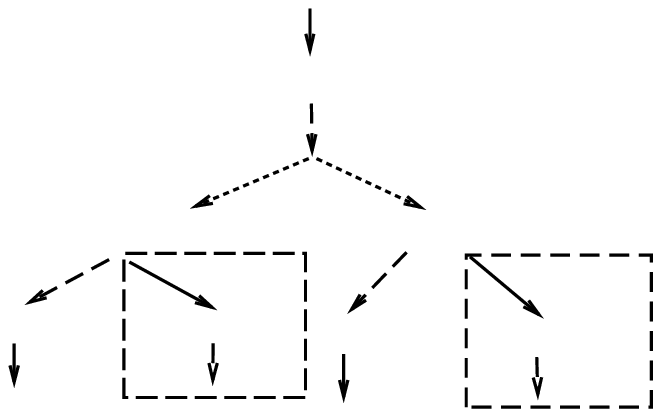}}
  \put(780.29,866.75){\fontsize{8.53}{10.24}\selectfont \makebox(30.0, 60.0)[l]{$P$\strut}}
  \put(734.62,793.51){\fontsize{6.83}{8.19}\selectfont \makebox(48.0, 48.0)[l]{10\strut}}
  \put(337.54,670.11){\fontsize{8.53}{10.24}\selectfont \makebox(1050.0, 60.0)[l]{$P_1\equiv\PAR{\OUTPUT[c]{b_1,\infty_2}}{\PAR{ \OUTPUT[c]{b_2,\infty_3}}{\PREFIX{\INPUT[c]{x,6}}{\OUTPUT[c]{x,6}}}}$\strut}}
  \put(757.40,576.63){\fontsize{6.83}{8.19}\selectfont \makebox(24.0, 48.0)[l]{5\strut}}
  \put(163.06,355.26){\fontsize{8.53}{10.24}\selectfont \makebox(570.0, 60.0)[l]{$P_2\equiv\PAR{ \OUTPUT[c]{b_2,\infty_3}}{\OUTPUT[c]{b_1,6}}$\strut}}
  \put(801.75,352.00){\fontsize{8.53}{10.24}\selectfont \makebox(570.0, 60.0)[l]{$P_3\equiv\PAR{ \OUTPUT[c]{b_1,\infty_2}}{\OUTPUT[c]{b_2,6}}$\strut}}
  \put(40.00,173.18){\fontsize{8.53}{10.24}\selectfont \makebox(240.0, 60.0)[l]{$P_4\equiv\OUTPUT[c]{b_1,6}$\strut}}
  \put(414.26,171.03){\fontsize{8.53}{10.24}\selectfont \makebox(270.0, 60.0)[l]{$P_5\equiv\OUTPUT[c]{b_2,\infty_3}$\strut}}
  \put(615.30,510.71){\fontsize{6.83}{8.19}\selectfont \makebox(72.0, 48.0)[l]{0.4\strut}}
  \put(922.11,500.76){\fontsize{6.83}{8.19}\selectfont \makebox(72.0, 48.0)[l]{0.6\strut}}
  \put(274.04,295.49){\fontsize{6.83}{8.19}\selectfont \makebox(24.0, 48.0)[l]{3\strut}}
  \put(528.88,295.61){\fontsize{6.83}{8.19}\selectfont \makebox(24.0, 48.0)[l]{6\strut}}
  \put(152.28,106.39){\fontsize{6.83}{8.19}\selectfont \makebox(24.0, 48.0)[l]{6\strut}}
  \put(552.46,98.57){\fontsize{6.83}{8.19}\selectfont \makebox(24.0, 48.0)[l]{3\strut}}
  \put(878.85,284.57){\fontsize{6.83}{8.19}\selectfont \makebox(24.0, 48.0)[l]{2\strut}}
  \put(1215.86,275.81){\fontsize{6.83}{8.19}\selectfont \makebox(24.0, 48.0)[l]{6\strut}}
  \put(1131.20,158.22){\fontsize{8.53}{10.24}\selectfont \makebox(270.0, 60.0)[l]{$P_7\equiv\OUTPUT[c]{b_1,\infty_2}$\strut}}
  \put(788.19,160.04){\fontsize{8.53}{10.24}\selectfont \makebox(240.0, 60.0)[l]{$P_6\equiv\OUTPUT[c]{b_2,6}$\strut}}
  \put(890.19,97.08){\fontsize{6.83}{8.19}\selectfont \makebox(24.0, 48.0)[l]{6\strut}}
  \put(1211.27,81.73){\fontsize{6.83}{8.19}\selectfont \makebox(24.0, 48.0)[l]{2\strut}}
  \end{picture}%
\caption{\label{fig:immediate transition}Example with Immediate Transitions}
\end{minipage}
\begin{minipage}[t]{0.4\textwidth}
\centering
  \setlength{\unitlength}{0.05 mm}%
  \begin{picture}(509.5, 532.0)(0,0)
  \put(0,0){\includegraphics{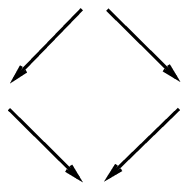}}
  \put(229.72,440.45){\fontsize{7.11}{8.54}\selectfont \makebox(25.0, 50.0)[l]{$P$\strut}}
  \put(164.56,334.62){\fontsize{5.69}{6.83}\selectfont \makebox(20.0, 40.0)[l]{4\strut}}
  \put(40.00,241.79){\fontsize{7.11}{8.54}\selectfont \makebox(75.0, 50.0)[l]{$P_4$\strut}}
  \put(151.98,169.93){\fontsize{5.69}{6.83}\selectfont \makebox(20.0, 40.0)[l]{6\strut}}
  \put(317.55,169.86){\fontsize{5.69}{6.83}\selectfont \makebox(20.0, 40.0)[l]{6\strut}}
  \put(394.48,240.69){\fontsize{7.11}{8.54}\selectfont \makebox(75.0, 50.0)[l]{$P_6$\strut}}
  \put(305.56,338.80){\fontsize{5.69}{6.83}\selectfont \makebox(20.0, 40.0)[l]{6\strut}}
  \put(238.50,38.50){\fontsize{7.11}{8.54}\selectfont \makebox(25.0, 50.0)[l]{$0$\strut}}
  \end{picture}%
\caption{\label{fig:Immediate CTMC}CTMC corresponding to $P$}
\end{minipage}
\end{figure}

\begin{example}\label{ex:immediate transition}
Suppose $P \equiv
\PAR{\OUTPUT[a]{b,10}}{\PAR{\PREFIX{\INPUT[a]{x,1}}{\OUTPUT[c]{b_1,\infty_2}}}{\PAR{\PREFIX{\INPUT[a]{x,1}}{\OUTPUT[c]{b_2,\infty_3}}}{\PREFIX{\INPUT[c]{x,1}}{\OUTPUT[c]{x,6}}}}}$,
by the semantics in Table \ref{tab:inference rules} we can draw a
derivation tree as Fig.~\ref{ex:immediate transition}. We
 omit passive transitions like $\TRAN{}{0}{}$ here.

In Fig.~\ref{fig:immediate transition} nondeterministic choices
emerge, such as process
$\PAR{\OUTPUT[c]{b_2,\infty_3}}{\OUTPUT[c]{b_1,6}}$ can choose
either immediate transition $\TRANI{}{3}{}$ or Markovian transition
$\TRAN{}{6}{}$. But since the probability of the Markovian
transition being executed before the immediate one is zero, so the
transitions inside the dashed rectangle is impossible and should be
ignored.

In this example, $\DER{P} = \{P, P_1,P_2,P_3,P_4,P_5,P_7,0\}$,
$\DER{P} \cap \MATHITSF{IP} = \{P_1,P_2,P_3,P_5,P_7\}$, and $\DER{P}
\cap \MATHITSF{MP}=\{P,P_4,P_6,0\}$. To define the CTMC of $P$, we
only need to consider the processes in $\DER{P} \cap \MATHITSF{MP}$
and the corresponding CTMC of $P$ is shown in
Fig.~\ref{fig:Immediate CTMC} which is quite simple compared to the
derivation tree in Fig.~\ref{fig:immediate transition}.
\end{example}

In the second case of Equation~(\ref{eq:prob}), we require that $\DER{R}\cap\MATHITSF{MP}\neq\emptyset$, that is, there exists at least a Markovian process which is reachable from $R$. But sometimes it is also possible for one process reaching an immediate state from which no Markovian process can be reached, that is, we have immediate loop. We call states in $\MATHITSF{SP}=\{P\mid\DER{P}\subseteq \MATHITSF{IP}\}$ absorbing states and use a special process $\MATHITSF{Stuck}$ to denote them. Accordingly, the set of states of the CTMC
should be $(\DER{P} \cap \MATHITSF{MP}) \cup \{\MATHITSF{Stuck}\}$
and $\MATRIX(P,\MATHITSF{Stuck})$ where $P \in \MATHITSF{MP}$ can be
defined as follows:
\[\MATRIX(P,\MATHITSF{Stuck}) = \lambda - \sum_{Q \in \DER{P}\cap\MATHITSF{MP}}\MATRIX(P,Q) \quad \TRAN{P}{\lambda}{\PDA{P}}.\]

\section{Conclusions and Future Works}
In this paper we give a stochastic broadcast $\pi$ calculus which is
useful to model some server-client based systems where synchronization is governed by only one participant. Both Markovian transitions and immediate transitions are taken into account. A few examples are given to show the expressivity of the calculus which is enhanced a lot with immediate transitions. The
semantics is given by Labeled Transition System without relying on techniques such as multi relations,
proved transition systems, unique rate names, and so on. Each transition is labeled with rate or weight instead of action and the resulting distribution is over pairs of actions and processes instead of only processes. In this way, the underlying CTMC can be obtained naturally
even with existence of immediate transitions.

A number of further developments are possible. In the future we would like to provide semantics to some of the most representative stochastic process languages as \cite{de2009rate,bernardo2010uniform} and compare these different ways. Another possible extension is to support parameters, that is, we do not need to know the value of
each parameter at beginning. We can also reuse the model by
assigning parameters with different values and so on. Sometimes,
some CTMCs have special form, that is, product form which can be
solved efficiently \cite{hillston1999product,harrison2004reversed}. We try to answer whether the underlying CTMC of
a process is in product form or not by syntax-checking.

\bibliographystyle{eptcs}
\bibliography{refs}

\begin{thebibliography}{10}
\providecommand{\bibitemdeclare}[2]{}
\providecommand{\urlprefix}{Available at }
\providecommand{\url}[1]{\texttt{#1}}
\providecommand{\href}[2]{\texttt{#2}}
\providecommand{\urlalt}[2]{\href{#1}{#2}}
\providecommand{\doi}[1]{doi:\urlalt{http://dx.doi.org/#1}{#1}}
\providecommand{\bibinfo}[2]{#2}

\bibitemdeclare{book}{aldini2009process}
\bibitem{aldini2009process}
\bibinfo{author}{A.~Aldini}, \bibinfo{author}{M.~Bernardo} \&
  \bibinfo{author}{F.~Corradini} (\bibinfo{year}{2009}):
  \emph{\bibinfo{title}{{A process algebraic approach to software architecture
  design}}}.
\newblock \bibinfo{publisher}{Springer-Verlag New York Inc}.

\bibitemdeclare{article}{bergstra1984process}
\bibitem{bergstra1984process}
\bibinfo{author}{J.A. Bergstra} \& \bibinfo{author}{J.W. Klop}
  (\bibinfo{year}{1984}): \emph{\bibinfo{title}{{Process algebra for
  synchronous communication}}}.
\newblock {\sl \bibinfo{journal}{Information and Control}}
  \bibinfo{volume}{60}(\bibinfo{number}{1-3}), pp. \bibinfo{pages}{109--137},
  \doi{10.1016/S0019-9958(84)80025-X}.

\bibitemdeclare{conference}{bernardo2007weak}
\bibitem{bernardo2007weak}
\bibinfo{author}{M.~Bernardo} \& \bibinfo{author}{A.~Aldini}
  (\bibinfo{year}{2007}): \emph{\bibinfo{title}{{Weak Markovian bisimilarity:
  abstracting from prioritized/weighted internal immediate actions}}}.
\newblock In: {\sl \bibinfo{booktitle}{Theoretical Computer Science:
  Proceedings of the 10th Italian Conference on ICTCS'07}}.
  \bibinfo{organization}{World Scientific Pub Co Inc}, p.~\bibinfo{pages}{39},
  \doi{10.1142/9789812770998(0)0008}.

\bibitemdeclare{article}{bernardo2010uniform}
\bibitem{bernardo2010uniform}
\bibinfo{author}{M.~Bernardo}, \bibinfo{author}{R.~De~Nicola} \&
  \bibinfo{author}{M.~Loreti} (\bibinfo{year}{2010}):
  \emph{\bibinfo{title}{{Uniform Labeled Transition Systems for
  Nondeterministic, Probabilistic, and Stochastic Processes}}}.
\newblock {\sl \bibinfo{journal}{Trustworthly Global Computing}} , pp.
  \bibinfo{pages}{35--56}\doi{10.1007/978-3-642-15640-3(0)3}.

\bibitemdeclare{article}{bernardo1994mpa}
\bibitem{bernardo1994mpa}
\bibinfo{author}{M.~Bernardo}, \bibinfo{author}{L.~Donatiello} \&
  \bibinfo{author}{R.~Gorrieri} (\bibinfo{year}{1994}):
  \emph{\bibinfo{title}{{MPA: a stochastic process algebra}}}.
\newblock {\sl \bibinfo{journal}{University of Bologna}} .

\bibitemdeclare{article}{bernardo1998tutorial}
\bibitem{bernardo1998tutorial}
\bibinfo{author}{M.~Bernardo} \& \bibinfo{author}{R.~Gorrieri}
  (\bibinfo{year}{1998}): \emph{\bibinfo{title}{{A tutorial on EMPA: A theory
  of concurrent processes with nondeterminism, priorities, probabilities and
  time}}}.
\newblock {\sl \bibinfo{journal}{Theoretical Computer Science}}
  \bibinfo{volume}{202}(\bibinfo{number}{1-2}), pp. \bibinfo{pages}{1--54},
  \doi{10.1016/S0304-3975(97)00127-8}.

\bibitemdeclare{article}{buzen1973computational}
\bibitem{buzen1973computational}
\bibinfo{author}{J.P. Buzen} (\bibinfo{year}{1973}):
  \emph{\bibinfo{title}{{Computational Algorithms for Closed Queueing Networks
  with Exponential Servers}}}.
\newblock {\sl \bibinfo{journal}{Communications of the ACM}}
  \bibinfo{volume}{16}(\bibinfo{number}{9}), pp. \bibinfo{pages}{527--531},
  \doi{10.1145/362342.362345}.

\bibitemdeclare{article}{de2007model}
\bibitem{de2007model}
\bibinfo{author}{R.~De~Nicola}, \bibinfo{author}{J.P. Katoen},
  \bibinfo{author}{D.~Latella}, \bibinfo{author}{M.~Loreti} \&
  \bibinfo{author}{M.~Massink} (\bibinfo{year}{2007}):
  \emph{\bibinfo{title}{{Model checking mobile stochastic logic}}}.
\newblock {\sl \bibinfo{journal}{Theoretical Computer Science}}
  \bibinfo{volume}{382}(\bibinfo{number}{1}), pp. \bibinfo{pages}{42--70},
  \doi{10.1145/362342.362345}.

\bibitemdeclare{article}{de2009rate}
\bibitem{de2009rate}
\bibinfo{author}{R.~De~Nicola}, \bibinfo{author}{D.~Latella},
  \bibinfo{author}{M.~Loreti} \& \bibinfo{author}{M.~Massink}
  (\bibinfo{year}{2009}): \emph{\bibinfo{title}{{Rate-based transition systems
  for stochastic process calculi}}}.
\newblock {\sl \bibinfo{journal}{Automata, Languages and Programming}} , pp.
  \bibinfo{pages}{435--446}\doi{10.1007/978-3-642-02930-1(0)36}.

\bibitemdeclare{article}{van1988jackson}
\bibitem{van1988jackson}
\bibinfo{author}{N.M. van Dijk} (\bibinfo{year}{1988}):
  \emph{\bibinfo{title}{{On Jackson's product form with jump-over blocking}}}.
\newblock {\sl \bibinfo{journal}{Operations Research Letters}}
  \bibinfo{volume}{7}(\bibinfo{number}{5}), pp. \bibinfo{pages}{233--235},
  \doi{10.1016/0167-6377(88)90037-5}.

\bibitemdeclare{article}{gotz1993multiprocessor}
\bibitem{gotz1993multiprocessor}
\bibinfo{author}{N.~G\"{o}tz}, \bibinfo{author}{U.~Herzog} \&
  \bibinfo{author}{M.~Rettelbach} (\bibinfo{year}{1993}):
  \emph{\bibinfo{title}{{Multiprocessor and distributed system design: The
  integration of functional specification and performance analysis using
  stochastic process algebras}}}.
\newblock {\sl \bibinfo{journal}{Performance evaluation of computer and
  communication systems}} , pp.
  \bibinfo{pages}{121--146}\doi{10.1007/BFb0013851}.

\bibitemdeclare{article}{harrison2004reversed}
\bibitem{harrison2004reversed}
\bibinfo{author}{P.G. Harrison} (\bibinfo{year}{2004}):
  \emph{\bibinfo{title}{{Reversed processes, product forms and a non-product
  form}}}.
\newblock {\sl \bibinfo{journal}{Linear Algebra and Its Applications}}
  \bibinfo{volume}{386}, pp. \bibinfo{pages}{359--381},
  \doi{10.1016/j.laa.2004.02.020}.

\bibitemdeclare{book}{hermanns2002interactive}
\bibitem{hermanns2002interactive}
\bibinfo{author}{H.~Hermanns} (\bibinfo{year}{2002}):
  \emph{\bibinfo{title}{{Interactive markov chains}}}.
\newblock \bibinfo{publisher}{Springer}, \doi{10.1007/3-540-45804-2(0)5}.

\bibitemdeclare{article}{hermanns1995formal}
\bibitem{hermanns1995formal}
\bibinfo{author}{H.~Hermanns}, \bibinfo{author}{M.~Rettelbach} \&
  \bibinfo{author}{T.~Weiss} (\bibinfo{year}{1995}):
  \emph{\bibinfo{title}{{Formal characterisation of immediate actions in SPA
  with nondeterministic branching}}}.
\newblock {\sl \bibinfo{journal}{The Computer Journal}}
  \bibinfo{volume}{38}(\bibinfo{number}{7}), p. \bibinfo{pages}{530},
  \doi{10.1093/comjnl/38.7.530}.

\bibitemdeclare{book}{hillston1996compositional}
\bibitem{hillston1996compositional}
\bibinfo{author}{J.~Hillston} (\bibinfo{year}{1996}): \emph{\bibinfo{title}{{A
  compositional approach to performance modelling}}}.
\newblock \bibinfo{publisher}{Cambridge University Press}.

\bibitemdeclare{article}{hillston1999product}
\bibitem{hillston1999product}
\bibinfo{author}{J.~Hillston} \& \bibinfo{author}{N.~Thomas}
  (\bibinfo{year}{1999}): \emph{\bibinfo{title}{{Product form solution for a
  class of PEPA models}}}.
\newblock {\sl \bibinfo{journal}{Performance Evaluation}}
  \bibinfo{volume}{35}(\bibinfo{number}{3-4}), pp. \bibinfo{pages}{171--192},
  \doi{10.1016/S0166-5316(99)00005-X}.

\bibitemdeclare{article}{hoare1978communicating}
\bibitem{hoare1978communicating}
\bibinfo{author}{C.A.R. Hoare} (\bibinfo{year}{1978}):
  \emph{\bibinfo{title}{{Communicating sequential processes}}}.
\newblock {\sl \bibinfo{journal}{Communications of the ACM}}
  \bibinfo{volume}{21}(\bibinfo{number}{8}), p. \bibinfo{pages}{677},
  \doi{10.1145/357980.358021}.

\bibitemdeclare{article}{lee1989approximate}
\bibitem{lee1989approximate}
\bibinfo{author}{H.S. Lee} \& \bibinfo{author}{S.M. Pollock}
  (\bibinfo{year}{1989}): \emph{\bibinfo{title}{{Approximate analysis for the
  merge configuration of an open queueing network with blocking}}}.
\newblock {\sl \bibinfo{journal}{IIE transactions}}
  \bibinfo{volume}{21}(\bibinfo{number}{2}), pp. \bibinfo{pages}{122--129},
  \doi{10.1080/07408178908966215}.

\bibitemdeclare{book}{milner1989communication}
\bibitem{milner1989communication}
\bibinfo{author}{R.~Milner} (\bibinfo{year}{1989}):
  \emph{\bibinfo{title}{{Communication and concurrency}}}.
\newblock \bibinfo{publisher}{Prentice Hall International Series in Computer
  Science}.

\bibitemdeclare{article}{priami1995stochastic}
\bibitem{priami1995stochastic}
\bibinfo{author}{C.~Priami} (\bibinfo{year}{1995}):
  \emph{\bibinfo{title}{{Stochastic $\pi$-calculus}}}.
\newblock {\sl \bibinfo{journal}{The Computer Journal}}
  \bibinfo{volume}{38}(\bibinfo{number}{7}), p. \bibinfo{pages}{578}.

\bibitemdeclare{article}{vigliotti2006stochastic}
\bibitem{vigliotti2006stochastic}
\bibinfo{author}{M.G. Vigliotti} \& \bibinfo{author}{P.G. Harrison}
  (\bibinfo{year}{2006}): \emph{\bibinfo{title}{{Stochastic ambient
  calculus}}}.
\newblock {\sl \bibinfo{journal}{Electronic Notes in Theoretical Computer
  Science}} \bibinfo{volume}{164}(\bibinfo{number}{3}), pp.
  \bibinfo{pages}{169--186}, \doi{10.1016/j.entcs.2006.07.018}.

\end{thebibliography}

\end{document}